\documentclass[11pt]{article}
\usepackage[utf8]{inputenc}
\usepackage[T1]{fontenc}
\usepackage{lmodern}
\usepackage[a4paper,margin=1in]{geometry}
\usepackage{amsmath,amssymb}
\usepackage{graphicx}
\usepackage{booktabs}
\usepackage{tabularx}
\usepackage[hidelinks]{hyperref}
\usepackage{caption}
\usepackage{orcidlink}
\graphicspath{{figuras/}}

\newcommand{\spm}{$s_{\pm}$}
\newcommand{\chio}{\ensuremath{\chi_0}}
\newcommand{\Ucrit}{\ensuremath{U_{\mathrm{crit}}}}
\newcommand{\EF}{\ensuremath{E_F}}
\newcommand{\hP}{\ensuremath{h_{\mathrm{Pn}}}}

\title{\bfseries NaFeP: first pairing prediction for an unmade 111 iron phosphide,\\
and the pnictogen-height rule inside one compound}
\author{Reinaldo In\'acio\,\orcidlink{0009-0000-6594-6014}\\
\small BITA (Inova Simples), S\~ao Paulo, Brazil \quad\texttt{reinaldo.inacio@bitatech.com.br}}
\date{15 September 2026}

\begin{document}
\maketitle

\begin{abstract}
\noindent\emph{Companion/application paper to arXiv:2609.10614. Where that paper describes and
calibrates the judge and shows that generation re-finds what is known, this paper takes the next
step it pointed to: it applies the judge, end to end, to the single compound that survived the
judge's own mechanism-and-stability screen, and issues a falsifiable gap-symmetry prediction.
Nothing here is a measurement, and no $T_c$ in Kelvin is asserted.}

\medskip
The prior work~\cite{ref1} established a physics judge that classifies the gap symmetry of a
superconductor candidate from its structure and showed, over three generation audits, that
generation in the wire-relevant families re-finds the known canon: geometry plus $d$-count is
necessary but not sufficient, and judgment---not generation---is the bottleneck. Here we act on that
conclusion. Starting from 58 iron-pnictide compositions in the electronic window of BaFe$_2$As$_2$
that are absent from the Materials Project, a zero-cost literature-and-stability screen leaves
exactly one compound with no experimental report as a bulk phase and no band-structure,
Fermi-surface, magnetic-order or pairing calculation (the high-throughput databases OQMD and
Alexandria record its hull position, a metallic gap and a zero moment, nothing more): NaFeP (111-type, space
group $P4/nmm$, Fe$^{2+}$ $d^6$, phosphorus ligand---no arsenic). The judge, refined since~\cite{ref1}
with an explicit Stoner gate and a frame-independent symmetry reader (Sec.~\ref{sec:judge}), returns
for NaFeP a paramagnetic ground state in DFT (ferromagnetic and N\'eel starts collapse; the stripe
order of the pnictides survives only as a $0.4\,\mu_B$ residue $0.4$~meV/Fe below the non-magnetic
state) and an \textbf{\spm{}} gap: a clean sign reversal between the hole cylinders at $\Gamma$ and the
electron sheets at the zone corner, at the nesting vector $(\tfrac{1}{2},\tfrac{1}{2},\tfrac{1}{2})$, in every
basis and cell we tried. Whether that gap is nodeless is where the instrument is tested. The
phosphorus height above the iron plane (1.16~\AA{} in the PBE-relaxed cell) lies below the empirical
threshold of $\approx1.33$~\AA{} that separates nodal from nodeless iron pnictides, so the
established structural rule calls NaFeP nodal. The sealed first verdict, on a metal-only ($d$)
Wannier basis, said nodeless with a modest margin; a $d+p$ basis that passes the same gates, built to
test that reading, puts the gap on the inner hole cylinder at the node threshold
($\min|g|/\max|g| = 0.07$--$0.10$), and a scan of the phosphorus height reproduces, inside one
compound, the ingredients of the height rule: a third hole pocket at 1.25~\AA, a switch of the
nesting vector to $(\tfrac{1}{2},\tfrac{1}{2},0)$, a monotonic approach to the Stoner instability, and a
near-degeneracy of \spm{} and $d$-wave channels beyond 1.33~\AA. The prediction we stand behind
is therefore the $d+p$ reading: an \spm{} superconductor with a near-nodal inner hole sheet, on the
LiFeP side of the height rule, with the instrument's disagreement between its two bases reported
rather than resolved by choice. Conditional on the nodeless reading, the manufacturability funnel returns a
wire-candidate profile (grain-boundary class~B, anisotropy $\gamma = 4.2$, coherence length
$\approx15$~nm at an assumed $T_c$ of 20~K, arsenic-free); the near-nodal reading erodes that
profile. No $T_c$ is asserted. NaFeP is reported as a synthesis target for which the discriminating
measurement, nodal or nodeless, is stated in advance. Total cloud cost of the NaFeP
characterization, including the height scan: about US\$\,25.
\end{abstract}

\section{Introduction}
The companion paper~\cite{ref1} framed the problem and built the instrument. Generation of candidate
crystal structures is cheap and abundant; the field's own referees have shown that generated
novelty is largely rediscovery~\cite{ref2,ref3}; and the public benchmark that ranks generators
measures thermodynamic stability, not whether a candidate can pair, with which symmetry, and whether
it can be drawn into wire. The gap symmetry decides the route to wire before any furnace is lit: a
$d$-wave gap makes the critical current collapse at grain boundaries and forces biaxially textured,
coated conductors~\cite{hilgenkamp}, while an \spm{} gap---sign-changing between Fermi sheets but
nodeless within them---tolerates grain boundaries~\cite{katase} and admits round wire. A judge that
classifies this, and says no when the physics is absent, is the layer between generation and
synthesis.

That paper audited generation by three independent routes and concluded that, in the wire
families, there is no free new target: a descriptor sweep re-found the canon, 1{,}248 MatterGen
structures yielded zero candidates that were new, stable and carried a pairing motif, and mechanism
filters over 47{,}893 compounds re-derived the community's known analogies. The natural question it
left open is the one this paper answers: take the residue of such a screen---a compound that is
genuinely unreported, stable, and carries the motif---and run the judge on it. What does the
instrument say about a candidate that no one has made?

We do this for NaFeP. The point of this paper is not a new instrument (the instrument is~\cite{ref1},
here only refined) and not a discovery in the laboratory sense (nothing is synthesized). It is the
first concrete output of the judge applied to an unreported candidate: a specific, falsifiable
prediction---NaFeP is an \spm{}, arsenic-free, ambient-pressure superconductor whose gap is
near-nodal on the inner hole sheet in the metal-plus-ligand basis that is our production model, and
nodeless in the metal-only basis of the sealed first verdict---offered as a
synthesis target, with the confidence and the limits of the instrument stated plainly, and with
the one structural fact that decides between the two readings (Sec.~\ref{sec:height}) stated first.
Concretely, this paper adds to the record of NaFeP, none of which exists in any database or paper
we could find (Sec.~\ref{sec:select}; Table~\ref{tab:s1} in the Appendix): (i) its first band-resolved
electronic structure (bands, Fermi surface, orbital character---the OQMD and Alexandria entries
carry a band gap of zero, a density of states at \EF{} and a zero moment, and no bands); (ii) the
first test of the antiferromagnetic orders of the family (N\'eel and stripe) against the
paramagnetic state; (iii) its first pairing verdict, sealed and dated, and the revision of that
verdict by a second basis; (iv) its first placement on the pnictogen-height axis, and a scan of
that axis inside the compound; and (v) a manufacturability profile conditional on (iii).

\section[The judge, and what was refined since the companion paper]{The judge, and what was refined since~\cite{ref1}}\label{sec:judge}
The pipeline is that of~\cite{ref1}: DFT on the cell (GGA-PBE, PAW; for NaFeP, as for
BaFe$_2$As$_2$ in~\cite{ref1}, no Hubbard $U$ enters the self-consistent step---the interaction
enters only through the RPA vertex); a gated Wannier tight-binding model (band-tracking validator R5
and a filling gate that requires the model to reproduce the DFT occupation of every band crossing
\EF); the full rank-4 bare susceptibility \chio{} on a $24^3$ mesh with no energy
window~\cite{graser}; the linearized gap equation in the RPA with a Kanamori interaction on the $d$
blocks ($J/U = 0.15$), reported at $\alpha = U/\Ucrit = 0.90, 0.95, 0.97$ at $T = 130$~K; a symmetry
classifier; and a wire funnel F0--F5. The reader is referred to~\cite{ref1} for the full chain, its
gates, and its pre-registration protocol. Two refinements were made since~\cite{ref1}, both dated,
both introduced after a result contradicted an expectation, and both used here.

\paragraph{An explicit Stoner gate (provisional, and basis-dependent).} In~\cite{ref1} the
negative control SrRu$_2$As$_2$ was separated from the positive BaFe$_2$As$_2$ by its critical
interaction \Ucrit{} (4.17 vs 0.97~eV) rather than by symmetry, and the paper noted that
comparisons between materials must be made at absolute $U$. Nb had returned a null verdict at the
N1 gate of~\cite{ref1} (no selected wave vector, on two bases); on 14 September 2026 it was
nonetheless taken through the gap equation, on its $s+d$ basis with the interaction on the $d$
block, as a test of the reader, and a reader that used the gap shape alone mislabelled it ``$d$'',
although its $\Ucrit = 3.65$~eV sits next to the ruthenide's. That value is the rank-4 Kanamori
quantity used everywhere in this paper, $1/\max_q \lambda[\chio(q)\,U]$ at $U = 1$,
$J/U = 0.15$, on the complete \chio{} of the $s+d$ model; it was recomputed from the stored
susceptibility for this revision (3.648~eV at $q^* = (0.42, 0.33, -0.08)$) and is not the maximum
of the scalar trace of \chio{}. We therefore made the correlation axis a coded gate-0: a material
with \Ucrit{} above a threshold is classified null/negative---far from the Stoner
instability---before any symmetry analysis. The threshold, 2.5~eV, is a round number in the gap
between the largest \Ucrit{} among the calibration superconductors (YBa$_2$Cu$_3$O$_7$, 1.32~eV)
and the smallest among the conventional and negative materials (Nb, 3.65~eV), and coincides with
their midpoint (2.49~eV). Its chronology is stated plainly (Table~\ref{tab:s2} in the Appendix):
the sealed NaFeP verdict, with $\Ucrit = 1.72$~eV, was computed at 12:52 local time on 14
September; the Nb value at 14:16; the 2.5~eV gate was proposed at 14:52 and coded at 15:01 of the
same day. The threshold was therefore fixed with the candidate's \Ucrit{} already in hand, not
blind to it. What protects the verdict is not the choice of 2.5~eV but the width of the gap
between the candidate and the nearest control, and that width depends on which NaFeP value is
entered: with the $d$-only value (1.72~eV) any threshold between 1.72 and 3.65~eV gives the same
seven labels, but with the official $d+p$ prediction (2.37~eV on the relaxed cell, 2.47~eV on the
sealed CHGNet cell) the protected range is only 2.47--3.65~eV, and any threshold between 1.72 and
2.47~eV would label the official candidate null/negative. The range is narrow at its lower end
precisely because the candidate sits 0.03~eV below the gate on the sealed cell. The gate is also
basis-dependent, because
\Ucrit{} is defined on the $d$ block of models with different bases (Table~\ref{tab:stoner}):
NaFeP sits at 1.68~eV in its $d$-only model and at 2.37~eV in its $d+p$ model, the latter within
0.13~eV of the threshold, and on the CHGNet cell the $d+p$ value (2.47~eV) is within 0.03~eV of
it. Until the calibration materials have been re-run on a common basis (a $d+p$ BaFe$_2$As$_2$
and a $d+p$ Nb are the planned controls) the gate is provisional: it separates the present suite
cleanly, and it makes explicit what~\cite{ref1} argued, that the correlation axis, not the
symmetry label alone, separates a spin-fluctuation superconductor from a conventional metal, but
its number should not be read as a physical constant.

\begin{table}[htbp]\centering\small
\begin{tabular}{llcl}
\toprule
material & Wannier basis ($U$ on the $d$ block) & \Ucrit{} (eV) & vs.\ gate 2.5~eV\\
\midrule
BaFe$_2$As$_2$ & Fe-$d$ only (20 functions, 4\,Fe $\times$ 5\,$d$) & 0.97 & below\\
La$_{2-x}$Sr$_x$CuO$_4$ & Cu-$d$ $+$ O-$p$ (8 functions), $U$ on four Cu-$d$ in blocks of 2 & 1.09 & below\\
YBa$_2$Cu$_3$O$_7$ & planes $+$ chain model of~\cite{ref1}, $U$ on six functions in blocks of 1, 1, 4 & 1.32 & below\\
Nb & $s+d$ (6 functions), $U$ on the five $d$ & 3.65 & above\\
SrRu$_2$As$_2$ & Ru-$d$ $+$ As-$p$ (32 functions), $U$ on the twenty Ru-$d$ & 4.17 & above\\
\midrule
NaFeP, relaxed cell & Fe-$d$ only (10 functions) & 1.68 & below\\
NaFeP, CHGNet cell (sealed) & Fe-$d$ only (10 functions) & 1.72 & below\\
NaFeP, relaxed cell (official) & Fe-$d$ $+$ P-$p$ (16 functions), $U$ on the ten $d$ & 2.37 & below by 0.13\\
NaFeP, CHGNet cell & Fe-$d$ $+$ P-$p$ (16 functions), $U$ on the ten $d$ & 2.47 & below by 0.03\\
\bottomrule
\end{tabular}
\caption{The Stoner gate against the basis on which each \Ucrit{} is defined. The calibration rows
are the sealed models of~\cite{ref1} (Supplementary S1 there; tight-binding hashes 9e14dd29,
7ad3bd19, cb117265, 70e6a34a, b1a97f99), each \Ucrit{} being $1/\max_q\lambda[\chio(q)\,U]$ at
$U = 1$ on the stored susceptibility of that model. Values are not comparable across bases
(Sec.~\ref{sec:scan}); the gate is provisional until the calibration is repeated on a common
basis.}
\label{tab:stoner}
\end{table}

\paragraph{A frame-independent symmetry reader.} The naive symmetry-projection string of the gap
solver is blind to the solver's rotated frame: on axes rotated $45^\circ$, the $d_{x^2-y^2}$
($B_{1g}$) of the crystal appears as $B_{2g}$, and a string that checks only the rotated-$B_{1g}$
projection mislabels $d$-wave as ``\spm{}''. This mislabels four of the calibration materials (LSCO,
YBCO, MgB$_2$, SrRu$_2$As$_2$ all read ``\spm{}'' by the string). The class is therefore read from
the frame-independent physical signature already present in the gap data---the sign structure across
Fermi sheets (inter-sheet reversal $=$ \spm{}; sign change within a sheet with nodes $= d$; nodeless
ratio $\min|g|/\max|g| \geq 0.10$ per sheet) together with the Stoner gate---not from the projection
string. On the calibration suite this reader recovers the correct class for all materials
(Sec.~\ref{sec:cal}); the naive string is retained only as a diagnostic field. The verdicts
of~\cite{ref1} are not affected by this correction: there the class was read from the
irreducible-representation weights in the crystal frame ($A_{1g}$ weight 0.999 for
BaFe$_2$As$_2$, $B_{1g}$ weight 1.000 for LSCO), and the string corrected here was an auxiliary
field of the gap solver, not the classifier of~\cite{ref1}. The node threshold of the reader,
$\min|g|/\max|g| = 0.10$ per sheet, was fixed together with the reader on 14 September 2026 from
the sealed suite: sheets with symmetry nodes (both cuprates, MgB$_2$, one sheet of Nb and of the
ruthenide) give $< 0.001$, nodeless sheets give 0.28 (BaFe$_2$As$_2$) and 0.18 (the sealed NaFeP,
already known when the value was set); 0.10 was placed between the two groups. It is a reading
convention, not a physical scale, and its sensitivity is reported in Sec.~\ref{sec:scan}.

\subsection{Calibration and error basis}\label{sec:cal}
\begin{figure}[htbp]\centering
\includegraphics[width=0.72\linewidth]{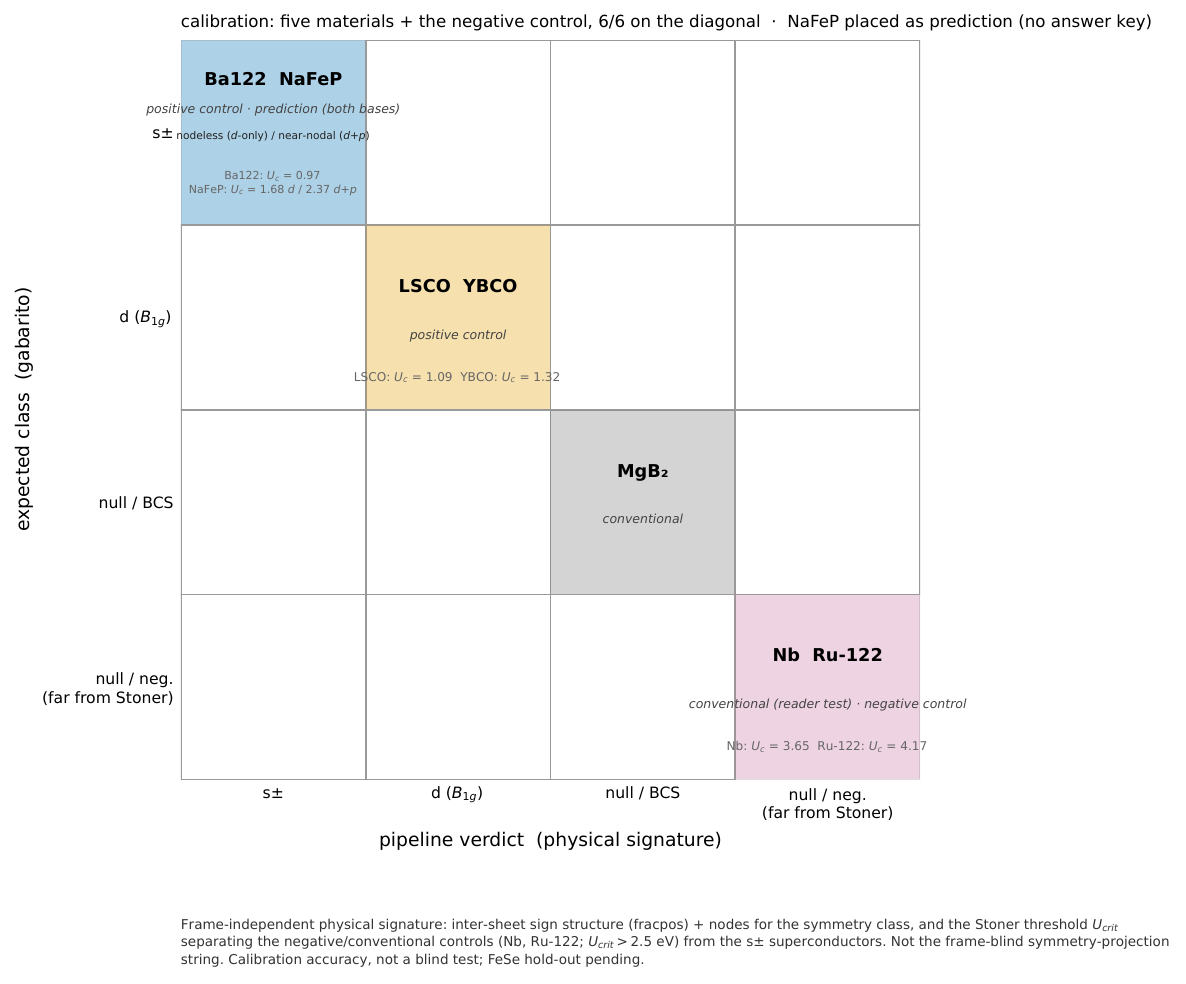}
\caption{Confusion matrix read by the frame-independent physical signature (inter-sheet sign
structure and nodes) together with the Stoner threshold \Ucrit{}. Five calibration materials and
the negative control on the diagonal; Nb and SrRu$_2$As$_2$ separate as null/negative
($\Ucrit>2.5$~eV). NaFeP is placed in the \spm{} cell as the prediction, not as a calibration
point, with both readings (nodeless in $d$-only, near-nodal in $d+p$) and both \Ucrit{} values.
Calibration accuracy, not a blind test; FeSe hold-out pending.}
\label{fig:confusion}
\end{figure}

Read by the physical signature plus the Stoner gate, the pipeline recovers the expected class for
the five calibration materials and the negative control with one instrument and no per-material
tuning, on a confusion matrix that is diagonal (Table~\ref{tab:cal}, Fig.~\ref{fig:confusion}). Nb$_3$Sn, the seventh material
of~\cite{ref1}, stops at the N1 gate of the classifier (no selected wave vector) and never reaches
the gap equation, so it has no \Ucrit{} and is not re-read here; its null verdict of~\cite{ref1}
stands. NaFeP appears in the table as the prediction, not as a calibration point.

\begin{table}[htbp]\centering\small
\begin{tabular}{llcc}
\toprule
expected class (answer key) & material & \Ucrit{} (eV) & pipeline verdict\\
\midrule
\spm{} & BaFe$_2$As$_2$ & 0.97 & \spm{}\\
$d$ ($B_{1g}$) & La$_{2-x}$Sr$_x$CuO$_4$ & 1.09 & $d$\\
$d$ ($B_{1g}$) & YBa$_2$Cu$_3$O$_7$ & 1.32 & $d$\\
null / BCS & MgB$_2$ & ---$^{a}$ & null\\
null / neg.\ (far from Stoner) & Nb & 3.65 & null/neg.\\
null / neg.\ (far from Stoner) & SrRu$_2$As$_2$ & 4.17 & null/neg.\\
\midrule
prediction (no answer key) & NaFeP, $d$-only basis (sealed) & 1.72 (1.68 relaxed) & \spm{} nodeless\\
prediction (no answer key) & \textbf{NaFeP, $d+p$ basis (official)} & \textbf{2.37} (2.47 CHGNet) & \textbf{\spm{}, near-nodal inner hole sheet}\\
\bottomrule
\end{tabular}
\caption{Verdicts by physical signature $+$ Stoner gate. Five calibration materials and the
negative control SrRu$_2$As$_2$ (answer keys pre-registered in~\cite{ref1}), and the prediction of
this paper in both bases; the $d+p$ row is the official prediction (Sec.~\ref{sec:scan}), and the
two \Ucrit{} values are on different scales (Table~\ref{tab:stoner}). $^{a}$MgB$_2$ is null by the
eigenvalue gate ($\lambda_1 = 0.013$ with complete \chio{}; no spin-fluctuation channel), before a
\Ucrit{} is defined.}
\label{tab:cal}
\end{table}

The two classes that determine wire viability---\spm{} (grain-boundary tolerant) and $d$
(grain-boundary sensitive)---are both demonstrated: BaFe$_2$As$_2$ as \spm{}, and both cuprates
(LSCO, YBCO) as $d$ against their forty-year experimental answer key. The conventional and negative
controls (MgB$_2$, Nb, SrRu$_2$As$_2$) return null, separated from the superconductors by the
Stoner gate. NaFeP falls in the \spm{} cell in both bases; the bases differ on the nodes, not on
the class.

We state the strength of this basis honestly, following~\cite{ref1}. \textbf{This is calibration
accuracy, not blind-test accuracy:} the corrections that produced the diagonal (the LSCO filling fix
of~\cite{ref1} and the two refinements above, the second of which was triggered by the Nb re-run)
were introduced after a result contradicted expectation, and no material was held out. A blind
hold-out with a pre-registered answer key---FeSe, its answer key sealed in the Supplementary
of~\cite{ref1}---is the next test and is not yet run. Five calibration materials and one negative
control are few. The
interaction $U$ is defined on the $d$ block of models with different bases (Fe-$d$ only for
BaFe$_2$As$_2$; Ru-$d$ $+$ As-$p$ for SrRu$_2$As$_2$), so the \Ucrit{} comparison across bases is a
limitation, and a $d+p$ re-run of BaFe$_2$As$_2$ is the planned control. The NaFeP prediction
inherits these limits: its \spm{} signature matches the positive control on the same chemistry and
basis, but the instrument's false-positive rate against a nodal alternative is bounded by a
calibration of five materials and one control, not by a blind test, and the calibration contains
no nodal iron pnictide.

\section{Candidate selection --- NaFeP as the survivor of a screen}\label{sec:select}
\begin{figure}[htbp]\centering
\includegraphics[width=0.95\linewidth]{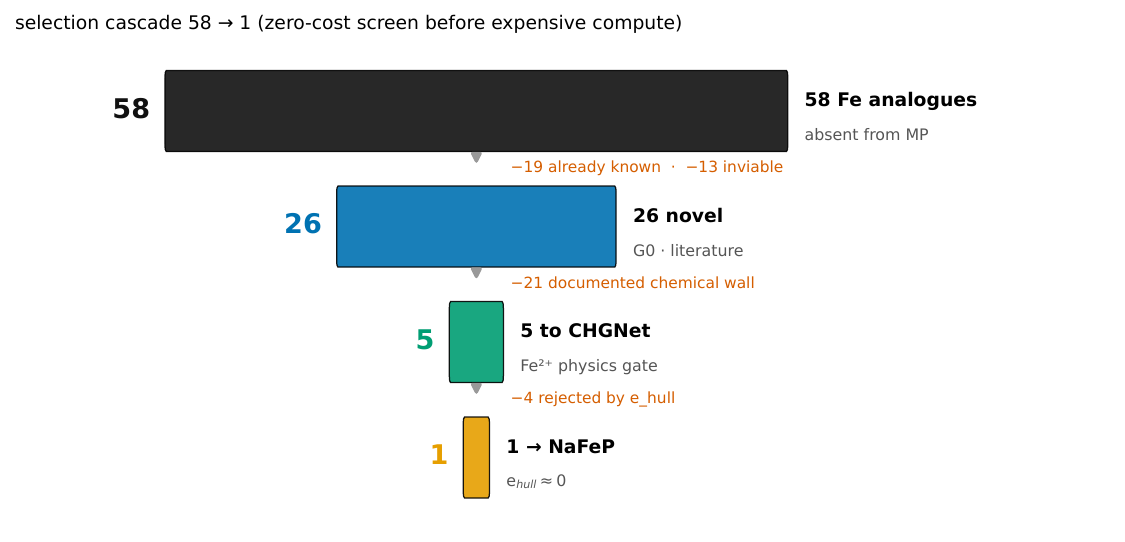}
\caption{Selection cascade: 58 Fe analogues absent from the Materials Project $\to$ 26 novel by
the literature gate $\to$ 5 without a documented chemical wall, sent to the CHGNet formation-energy
relaxation $\to$ 1 stable and unreported (NaFeP). The expensive pipeline is reserved for the single
survivor of a zero-cost screen.}
\label{fig:cascade}
\end{figure}

NaFeP was not chosen; it was the residue of an elimination (Fig.~\ref{fig:cascade}), and the
elimination is itself the method's discipline: the expensive pipeline is spent only on what survives
a zero-cost screen. We began from 58 iron-pnictide compositions in the electronic window of
BaFe$_2$As$_2$ (Fe$^{2+}$ $d^6$, square-net Fe, covalent ligand) that are absent from the Materials
Project~\cite{mp}. Each was subjected to a literature gate---a real search of web, arXiv, ICSD and
COD for prior synthesis, prior calculation, or a documented reason for non-formation. The gate
rejected 57 of 58 at zero compute cost:
\begin{itemize}
\item \textbf{19 already characterized} (7 synthesized, 12 computed)---the pipeline would be
redundant; among them, compounds ``absent from the Materials Project'' had hidden, e.g.\ LaFeSiO
(synthesized 2022, $T_c\approx10$~K) and CaKFe$_4$As$_4$ (the well-known 1144 superconductor),
confirming that absence from a database is not novelty~\cite{ref3}.
\item \textbf{13 inviable} by documented chemistry (Fe--Bi immiscibility; Sb--Sb square nets that
do not form; Fe$^{1+}$ charge-forbidden chalcogenide routes; charge-impossible 1144 chalcogenides).
\item Of the \textbf{26 nominally novel}, 21 carried a documented chemical wall of the above kind.
\item Of the 5 that reached a formation-energy relaxation (CHGNet~\cite{chgnet}, recomputed on the
candidate structure against the Materials Project hull of the chemical system), 4 were unstable or
metastable beyond the ambient threshold ($e_{\mathrm{hull}} = 0.19$--$0.23$~eV/atom).
\end{itemize}
One compound survived all gates: \textbf{NaFeP}, with $e_{\mathrm{hull}} = 0.000$~eV/atom. It is a
111-type structure (Cu$_2$Sb/PbFCl), the electronic analogue of LiFeP (nodal,
$T_c\approx6$~K)~\cite{deng2009,hashimoto} and LiFeAs (nodeless, $T_c\approx18$~K)~\cite{wang2008,tapp,hashimoto},
both non-magnetic, and of NaFeAs, which orders antiferromagnetically ($T_N\approx40$~K) and
superconducts at $T_c\approx9$--$23$~K depending on stoichiometry~\cite{parker,li2009}. Its ligand
is phosphorus, not arsenic---a manufacturability advantage in toxicity and processing.

\paragraph{What ``unreported'' means for NaFeP.} Precision matters here, because~\cite{ref1} argued
that absence from a database is not novelty. NaFeP is absent from the Materials Project. It is
\emph{present} in the Open Quantum Materials Database~\cite{oqmd} as a prototype-decorated DFT
entry (entry 1370778, $P4/nmm$, formation energy $-0.61$~eV/atom, on the OQMD convex hull; a
$Cmma$ variant of the same layer lies 0.4~meV/atom above it), i.e.\ its formation energy has been
computed in a high-throughput sweep, and that independent PBE hull agrees with our CHGNet hull. It
is also present in the Alexandria database~\cite{alexandria} (entry agm002166081, $P4/nmm$, on the
PBE and PBEsol hulls, formation energy $-0.48$~eV/atom in PBE), whose high-throughput
spin-polarized relaxation records a band gap of zero, a density of states at \EF{} of 2.18
states/eV per cell and a zero magnetic moment, and no band structure. It is absent from
JARVIS-DFT~\cite{jarvis} and from the Crystallography Open Database~\cite{cod}, and the only
$P4/nmm$ Na--Fe--P entries in AFLOW~\cite{aflow} are prototype decorations with sodium on the
square-net site and iron on the anion site (formation enthalpy $+0.11$~eV/atom), not the 111
compound; the ICSD itself was not accessible to us (Table~\ref{tab:s1} in the Appendix lists every
database, the query and the date). No band-resolved electronic-structure, magnetic-order or
pairing calculation of NaFeP exists in the literature we could find, and no bulk synthesis or
crystal-structure refinement is reported; a ``metastable ternary
tetragonal NaFeP'' is mentioned as a sodiation intermediate of FeP anodes in the sodium-ion battery
literature~\cite{yan2025}, without a structure determination. The claim of this paper is therefore
not that NaFeP has never been written down, but that no one has computed its bands, its Fermi
surface, its magnetic orders or its pairing, and no one has made it as a bulk phase. Two caveats on stability: the CHGNet hull
rests on a Na--Fe--P ternary space that is empty in the Materials Project (55 competing phases, all
binary or elemental), so $e_{\mathrm{hull}}\approx0$ means ``no decomposition driving force seen'',
not a deep margin; and the near-degenerate $Cmma$ entry in OQMD signals a polymorphism risk that
synthesis must resolve. Both prior mentions otherwise work in the candidate's favour: the OQMD hull
is an independent PBE confirmation of the stability that CHGNet found, and the battery observation
that sodium enters the FeP layer electrochemically suggests a low-temperature route (electrochemical
or ammonothermal-type intercalation of FeP) alongside conventional solid-state synthesis. None of
this contradicts the conclusion of~\cite{ref1} that generation re-finds the known: NaFeP was not
generated, it was enumerated as the sodium analogue of a known family and it already sits on two
high-throughput hulls; what is new here is the judgment applied to it, not the structure.

\section{Methods applied to NaFeP}\label{sec:methods}
\begin{figure}[htbp]\centering
\includegraphics[width=0.9\linewidth]{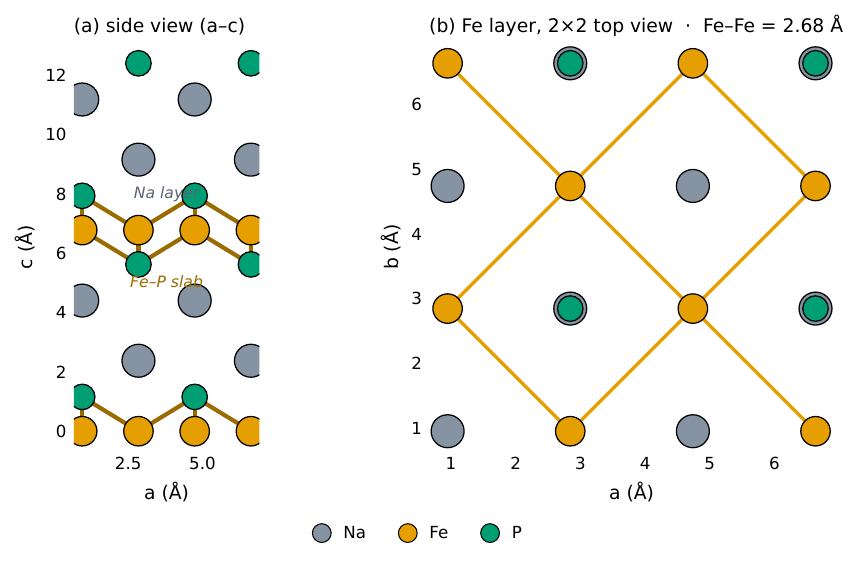}
\caption{Crystal structure of NaFeP ($P4/nmm$, PBE-relaxed cell). (a) Side view ($a$--$c$):
Fe--P slabs with Na layers intercalated between them. (b) Top view ($2\times2$): the square Fe
sublattice, Fe--Fe $=2.66$~\AA.}
\label{fig:struct}
\end{figure}

\paragraph{Structure.} The screen of Sec.~\ref{sec:select} delivered a CHGNet-relaxed cell
($a = b = 3.765$~\AA, $c = 7.050$~\AA, six atoms, P at $z = 0.1615$), on which the first, sealed
verdict of Sec.~\ref{sec:results} was computed. That cell was then fully relaxed in DFT (PBE,
variable cell and internal coordinates, BFGS, residual pressure $< 0.2$~kbar, 17 steps):
$a = b = 3.7908$~\AA, $c = 6.7779$~\AA, $V = 97.40$~\AA$^3$, P at $z = 0.1706$, Na at $z = 0.3504$
(Fig.~\ref{fig:struct}). The stacking axis contracts by 3.9\% relative to CHGNet and the in-plane
constant grows by 0.7\%; the phosphorus height moves from 1.139 to $\hP = 1.156$~\AA, the Fe--P bond
is 2.220~\AA{} and the P--Fe--P angles are $117.2^\circ$ (twice) and $105.7^\circ$ (four times), a
tetrahedron flattened well beyond that of LiFeP. The independent PBE relaxations in OQMD~\cite{oqmd}
($a = 3.786$~\AA, $c = 6.677$~\AA, $\hP = 1.148$~\AA) and in Alexandria~\cite{alexandria}
($a = 3.797$~\AA, $c = 6.698$~\AA, $\hP = 1.150$~\AA; PBEsol: 3.748, 6.653, 1.137~\AA) agree with
ours to 0.2\% in $a$, 1.5\% in $c$ and 0.7\% in \hP{}, which places the CHGNet $c$ axis, not the FeP
layer, as the quantity the machine learning potential had wrong. Every result below, unless labelled otherwise, is on the relaxed cell;
the CHGNet-cell verdict is kept as the dated original. Pseudopotentials were SSSP~1.3.0
precision~\cite{sssp} (Na, Fe, P), plane-wave cutoffs 100/1080~Ry, in Quantum ESPRESSO
7.4.1~\cite{qe}.

\paragraph{Magnetic state.} Four spin-polarized configurations were compared with the non-magnetic
state, each against a non-magnetic reference in the same cell and $k$-mesh: ferromagnetic (2-Fe cell),
N\'eel antiferromagnetic (2-Fe cell, $12\times12\times6$), and the stripe antiferromagnetic order of
the iron pnictides, wave vector $(\tfrac{1}{2},\tfrac{1}{2},0)$ of the 2-Fe cell, in the
$\sqrt2\times\sqrt2\times1$ supercell (4 Fe, 12 atoms, $8\times8\times6$). The ferromagnetic and
N\'eel starts collapse to zero moment ($\Delta E = +0.05$~meV/atom and $+0.15$~meV/Fe). The stripe
start keeps a residual moment of $0.40\,\mu_B$/Fe and lies $0.42$~meV/Fe \emph{below} the
non-magnetic state: an instability that is real in PBE but two orders of magnitude weaker than the
$\sim0.1$~eV/Fe and $\sim1.7\,\mu_B$ that the same functional gives LiFeAs, a material that does not
order~\cite{yin2011}, and below $k_BT$ at the 130~K of the susceptibility. On the CHGNet cell the
same four starts all collapse ($\Delta E = +0.15$~meV/Fe for both antiferromagnetic patterns,
$|m| \leq 0.003\,\mu_B$/Fe), so the residual stripe moment is a property of the shorter relaxed
$c$ axis, not of the FeP layer; the spin-polarized high-throughput relaxation of
Alexandria~\cite{alexandria}, from a ferromagnetic start, also ends at zero moment. Given that GGA
overestimates ordered moments across the family~\cite{mazin2008dft}, NaFeP is, within DFT, as
non-magnetic as an iron pnictide gets; the paramagnetic state that the RPA of~\cite{ref1} uses for
BaFe$_2$As$_2$ is here not an assumption but the DFT ground state to within a fraction of a meV. The
original protocol (antiferromagnetic cells only if the ferromagnetic moment exceeded
$0.3\,\mu_B$/Fe) was tightened after the first version of this paper; the stripe test was added on
14 September 2026, before any pairing calculation on the relaxed cell.

\begin{figure}[htbp]\centering
\includegraphics[width=0.95\linewidth]{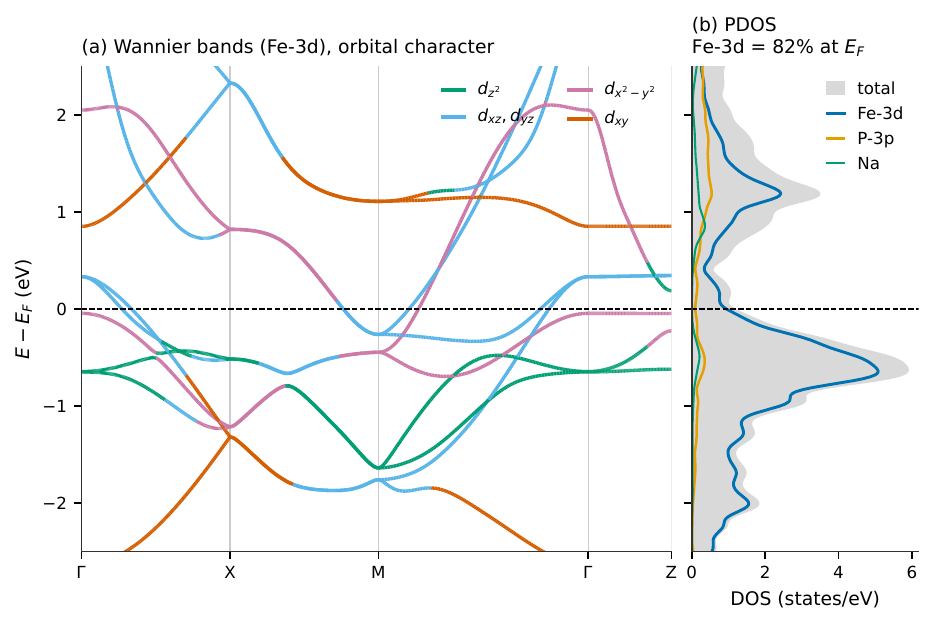}
\caption{(a) Wannier bands (10 Fe-3$d$ WF) along $\Gamma$--X--M--$\Gamma$--Z, colored by dominant
$d$-orbital character. (b) Species-projected density of states: Fe-3$d$ makes up 82\% of $N(\EF)$.}
\label{fig:bands}
\end{figure}

\paragraph{Electronic structure.} A non-self-consistent calculation on an explicit
$12\times12\times12$ crystal-coordinate $k$-mesh (1728 points) followed, with L\"owdin projections
retained. Four sheets cross \EF{}, all of essentially pure $d$ character: two hole cylinders around
$\Gamma$ and two electron sheets at the zone corner, with no third ($d_{xy}$) hole pocket---the
Fermi-surface topology expected for a pnictide with a low anion height (Sec.~\ref{sec:height}). The
Fe-3$d$ fraction of $N(\EF)$ is 82\% on the relaxed cell (P-3$p$ 8\%, Na 1\%; 80\% on the CHGNet
cell; Fig.~\ref{fig:bands}), which passes the
transferability rule of~\cite{ref1} for a metal-only basis ($\geq70\%$).

\begin{figure}[htbp]\centering
\includegraphics[width=0.95\linewidth]{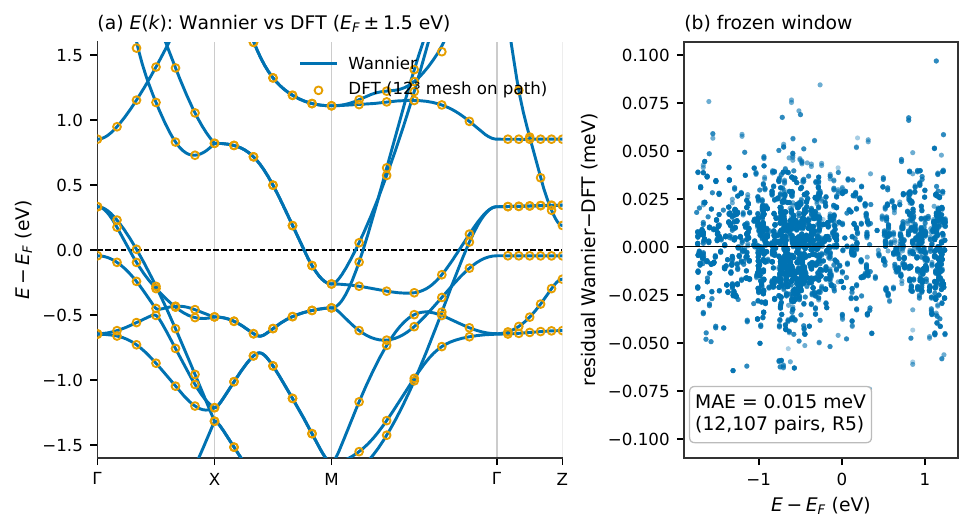}
\caption{R5 band-tracking validation of the $d$-only model on the relaxed cell (frozen window
$\EF-1.8$ to $\EF+1.5$~eV). (a) Wannier $E(k)$ (lines) over the DFT eigenvalues at the
$12^3$-mesh points on the path (circles), $\EF\pm1.5$~eV. (b) Residual over the frozen window
(12{,}107 pairs), MAE $=0.015$~meV.}
\label{fig:r5}
\end{figure}

\paragraph{Wannier Hamiltonian.} Two bases were built on each NSCF: A, Fe-$d$ $+$ P-$p$ (16
Wannier functions), and B, Fe-$d$ only (10 functions, 2\,Fe $\times$ 5\,$d$), the latter matching the
basis convention of the BaFe$_2$As$_2$ positive control (the ligand-$p$ tails enter the effective
$d$--$d$ hoppings of the downfolded model). Both reproduce the DFT bands inside the frozen window
$\EF\pm1.5$~eV to a mean absolute error of $0.010$--$0.015$~meV. On the CHGNet cell of the sealed
verdict, model A as first built (with a wide outer window) failed the band-count criterion of the
R5 validator (one state more than DFT inside the frozen window at 1270 of the 1728 $k$-points), so
the sealed verdict is on model B, which passes R5 on both criteria (Fig.~\ref{fig:r5}) and the
filling gate. Model A, rebuilt with an outer window taken from the DFT eigenvalues, passes both
gates on every cell and height of Sec.~\ref{sec:scan} and is the production basis of the official
prediction; the two bases, and what the gates said about each, are set out there.

\begin{figure}[htbp]\centering
\includegraphics[width=0.95\linewidth]{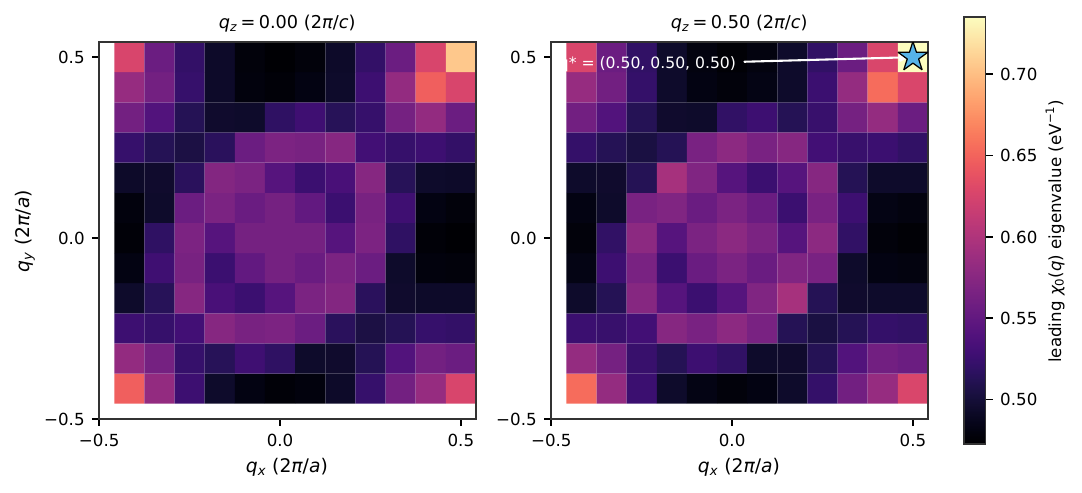}
\caption{Leading eigenvalue of the full-bandwidth bare susceptibility \chio{}$(q)$ (Graser/Kubo
standard~\cite{graser}, no energy window), planes $q_z=0$ and $q_z=0.5$. The nesting peak
$q^{*}=(0.5,0.5,0.5)$ (star) feeds the pairing vertex and coincides with the wavevector selected by
the gap equation; in BaFe$_2$As$_2$ the same instrument selects $(0.5,0.5,0)$, the stripe
wavevector~\cite{ref1}.}
\label{fig:chi0}
\end{figure}

\paragraph{Susceptibility and gap.} The full-bandwidth \chio{} (no energy window; the low-energy
restriction is applied by Fermi-surface projection in the gap equation; Fig.~\ref{fig:chi0}) and
the linearized gap equation on the Fermi surface were solved with complex-amplitude orbital weights
from the L\"owdin projections. A recomputation with checkpointing reproduced the sealed verdict
bit-for-bit ($\lambda_1$ identical to five decimals at all three $\alpha$), certifying the result.

\section{Results --- the gap symmetry of NaFeP}\label{sec:results}
\begin{figure}[htbp]\centering
\includegraphics[width=0.78\linewidth]{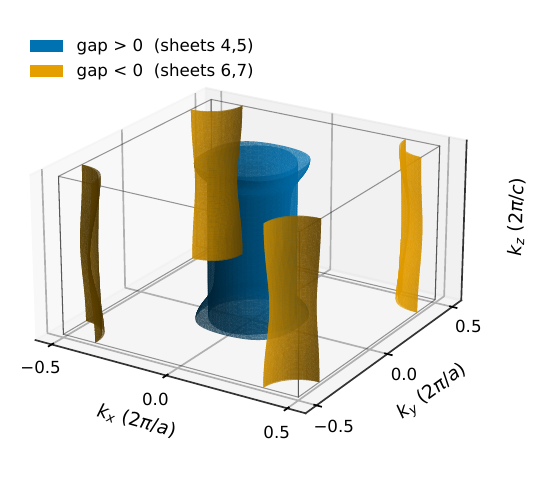}\\[4pt]
\includegraphics[width=0.95\linewidth]{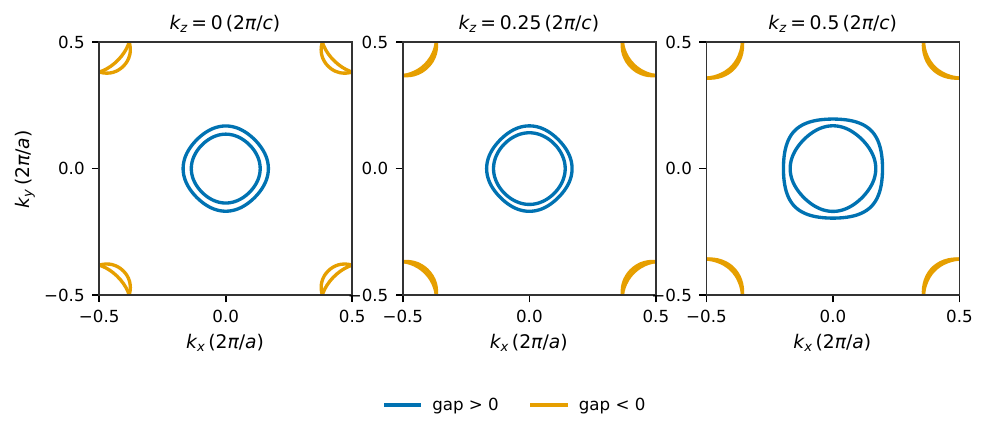}
\caption{Fermi surface of NaFeP colored by the sign of the leading gap eigenvector ($\alpha=0.90$,
$T=130$~K). Top: 3D in the Brillouin zone---the $\Gamma$-centered hole cylinders (sheets 4,5;
$\mathrm{gap}>0$, blue) reverse sign against the zone-corner electron sheets (sheets 6,7;
$\mathrm{gap}<0$, orange): the \spm{} signature. Bottom: $k_z=0,\,0.25,\,0.5$ cuts. Colorblind-safe
blue/orange.}
\label{fig:fs}
\end{figure}

The sealed verdict, computed on the CHGNet cell with the $d$-only basis on 14 September 2026 before
any of the robustness work of Sec.~\ref{sec:scan}, classifies NaFeP as \textbf{\spm{} nodeless}
across the coupling scan (Table~\ref{tab:gap}). Figures~\ref{fig:fs} and~\ref{fig:nodeless} show the
same quantities on the PBE-relaxed cell, where the $d$-only verdict is unchanged and the margin
larger. This section reports the sealed verdict as it was dated; the official prediction of the
paper, which revises its nodeless character, is stated in Sec.~\ref{sec:scan}.

\begin{table}[htbp]\centering\small
\begin{tabular}{ccccc}
\toprule
$\alpha = U/\Ucrit$ & $U$ (eV) & $\lambda_1$ & symmetry & $\min|g|/\max|g|$, sheets 4/5/6/7\\
\midrule
0.90 & 1.55 & 0.062 & \spm{} & 0.39 / 0.18 / 0.55 / 0.58\\
0.95 & 1.64 & 0.108 & \spm{} & 0.39 / 0.18 / 0.59 / 0.59\\
0.97 & 1.67 & 0.146 & \spm{} & 0.37 / 0.15 / 0.55 / 0.55\\
\bottomrule
\end{tabular}
\caption{The sealed verdict (CHGNet cell, $d$-only basis, 14 September 2026): leading pairing
eigenvalue, symmetry and nodeless ratio versus coupling ($\Ucrit = 1.72$~eV, $T = 130$~K, $24^3$
$k$-mesh, complete \chio{}).}
\label{tab:gap}
\end{table}

The \spm{} assignment rests on a clean sign reversal between Fermi-surface sheets: sheets 4 and 5
(hole) carry a positive gap, sheets 6 and 7 (electron) a negative gap (sign-fraction 1/1/0/0). Within
each sheet the gap is node-free ($\min|g|/\max|g| \geq 0.15$ on all sheets, above the 0.10 nodeless
threshold), the weakest sheet being the inner hole cylinder (sheet 5, 0.15--0.18). This is the
LiFeAs-type (nodeless) result on this basis, not the LiFeP-type (nodal)~\cite{hashimoto};
Sec.~\ref{sec:scan} shows that the ligand-explicit basis reverses this reading on the inner hole
sheet. The critical Hubbard coupling on this basis is $\Ucrit = 1.72$~eV (1.68~eV on the relaxed
cell; 2.37~eV in the $d+p$ basis, on its own scale), placing NaFeP as a moderately correlated 3$d$
system---near enough to
the Stoner instability to pair by spin fluctuations (below the 2.5~eV gate), but farther from it
than the positive control BaFe$_2$As$_2$ (0.97~eV), and well separated from the conventional and
negative materials (Nb 3.65, SrRu$_2$As$_2$ 4.17~eV). At the same absolute $U$ as BaFe$_2$As$_2$ at
its $\alpha=0.95$ (0.92~eV), NaFeP sits at $\alpha = 0.53$ of its own instability.

\paragraph{On the criterion for \spm{}.} The verdict is established by the inter-sheet sign
reversal, the defining signature of \spm{} pairing in the iron-based family~\cite{ref4,ref5,ref6},
not by a large $A_{1g}$ form-factor projection (which is small, $\leq0.013$, as expected for a state
that is uniform within sheets but reverses between them). We state this explicitly because a reader
may look for a dominant $\cos k_x+\cos k_y$ form factor; its absence is the correct behavior for
multi-band \spm{}.

\subsection{The pnictogen height, and where the prediction can fail}\label{sec:height}
The single structural parameter that the iron-pnictide literature ties most directly to the gap
structure is the anion height above the Fe plane, \hP{}. Spin-fluctuation theory
(Kuroki \emph{et al.}~\cite{kuroki2009}) predicts that lowering \hP{} removes the $d_{xy}$ hole
pocket, weakens the $(\pi,0)$ channel and lets nodes develop on the electron sheets, turning a
high-$T_c$ nodeless \spm{} state into a low-$T_c$ nodal one; the penetration-depth survey of
Hashimoto \emph{et al.}~\cite{hashimoto} across the 1111, 122 and 111 families places the empirical
switch at $\hP\approx1.33$~\AA, with LiFeP ($\hP\approx1.32$~\AA, $T_c = 6$~K) nodal and LiFeAs
($\hP\approx1.51$~\AA, $T_c = 18$~K) nodeless; and the $T_c$-versus-\hP{} curve of Mizuguchi
\emph{et al.}~\cite{mizuguchi} peaks at $\approx1.38$~\AA{} and falls to a few kelvin on the
phosphide side (LaFePO, $\hP\approx1.13$~\AA, $T_c\approx4$--$7$~K).

NaFeP has $\hP = 1.14$~\AA{} in the CHGNet cell and $1.15$~\AA{} in the OQMD PBE cell
(Table~\ref{tab:111}), below LiFeP and at the LaFePO end of the curve. PBE-type relaxations
underestimate pnictogen heights by up to $\sim0.1$~\AA~\cite{mazin2008dft}, so the true value may be
$\approx1.2$--$1.25$~\AA; it remains below the threshold. Our Fermi surface shows the consequence
(no $d_{xy}$ hole pocket, Sec.~\ref{sec:methods}), yet the RPA solution on the $d$-only model does
not develop nodes: the inner hole cylinder keeps $\min|g|/\max|g| = 0.15$--$0.18$, above the 0.10
line but closer to it than any other sheet. When the first version of this paper was written,
three readings were open: (i) the empirical rule, drawn from arsenides and from LiFeP, does not
extend this far down in \hP{} for a Na-spaced phosphide; (ii) the $d$-only model, which folds the
P-$p$ states into the $d$--$d$ hoppings, misses the orbital physics that produces the nodes; (iii)
the nodeless verdict is right and NaFeP is the first low-\hP{} nodeless pnictide. We then built the
$d+p$ model that could decide between them and scanned the height (Sec.~\ref{sec:scan}). The answer
is (ii): with phosphorus explicit, the inner hole cylinder goes to the node threshold at the
physical height, and the instrument agrees with the empirical rule. The \spm{} sign reversal between
hole and electron sheets was never at stake and is confirmed in both bases.

\begin{table}[htbp]\centering\small
\begin{tabular}{lccccll}
\toprule
compound & $a$ (\AA) & $c$ (\AA) & \hP{} (\AA) & magnetic order & $T_c$ (K) & gap\\
\midrule
LiFeP~\cite{deng2009,hashimoto} & 3.692 & 6.031 & $\approx$1.32 & none & 6 & nodal\\
LiFeAs~\cite{tapp,wang2008,hashimoto} & 3.791 & 6.364 & $\approx$1.51 & none & 18 & nodeless\\
NaFeAs~\cite{parker,li2009,liu2011} & 3.949 & 7.040 & $\approx$1.42 & stripe AFM, $T_N\approx40$~K & 9--23 & nodeless (Co-doped)\\
\midrule
NaFeP, PBE-relaxed cell (this work) & 3.791 & 6.778 & 1.16 & paramagnetic (PBE) & not predicted & \spm{}; nodeless ($d$) / near-nodal ($d+p$)\\
NaFeP, CHGNet cell (sealed original) & 3.765 & 7.050 & 1.14 & paramagnetic (PBE) & not predicted & \spm{}; nodeless ($d$) / near-nodal ($d+p$)\\
NaFeP, OQMD PBE cell~\cite{oqmd} & 3.786 & 6.677 & 1.15 & --- & --- & ---\\
NaFeP, Alexandria PBE cell~\cite{alexandria} & 3.797 & 6.698 & 1.15 & zero moment (PBE) & --- & ---\\
\bottomrule
\end{tabular}
\caption{NaFeP against the 111 family. Experimental structures and gap structures for the three
known members; four relaxed cells for NaFeP. The anion height places NaFeP below the nodal/nodeless
switch of~\cite{hashimoto}.}
\label{tab:111}
\end{table}

\subsection{Robustness: the relaxed cell, and the gap along the pnictogen-height axis}\label{sec:scan}
\begin{figure}[htbp]\centering
\includegraphics[width=0.98\linewidth]{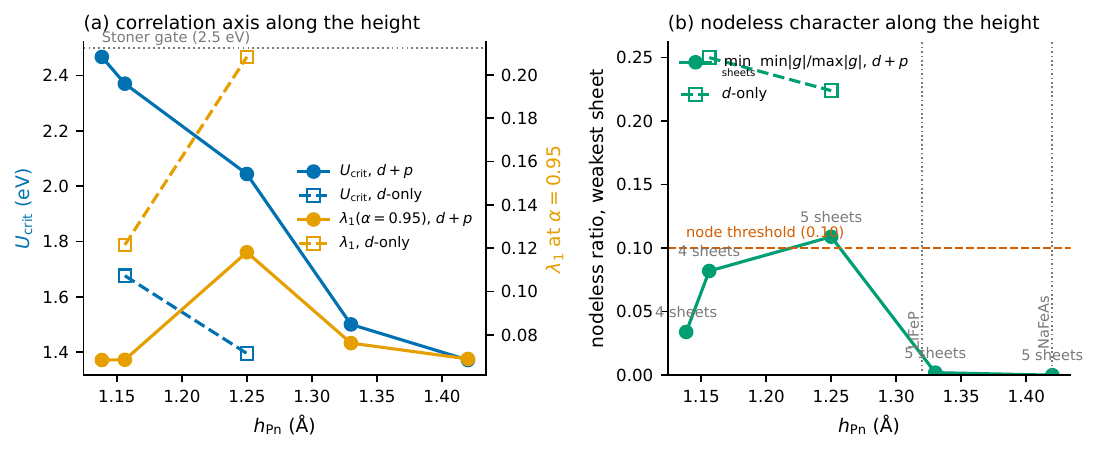}
\caption{The scan of Table~\ref{tab:scan}. (a) Critical interaction \Ucrit{} and leading eigenvalue
$\lambda_1$ ($\alpha = 0.95$) versus phosphorus height, $d+p$ (filled) and $d$-only (open) bases.
(b) Nodeless ratio of the weakest sheet; the node threshold of the reader is 0.10, and the heights of
LiFeP and NaFeAs are marked. In the $d+p$ basis the gap on the inner hole cylinder sits at the
threshold at the physical height and the leading channel loses its identity ($\lambda_2/\lambda_1
\to 1$) as the third hole pocket grows.}
\label{fig:scan}
\end{figure}

Because the whole prediction hangs on the nodeless character at a low anion height, we tested it
where it is weakest. The pipeline was re-run end to end on the PBE-relaxed cell of
Sec.~\ref{sec:methods}, and then on the same cell with the phosphorus displaced to
$\hP = 1.25$, $1.33$ and $1.42$~\AA{} (all other coordinates fixed; the last value is the height of
NaFeAs), the numerical experiment with which Kuroki \emph{et al.}~\cite{kuroki2009} first exposed the
height switch. Each point is a full DFT $\to$ Wannier $\to$ \chio{} $\to$ gap chain with the same
gates and the same interaction rule (Table~\ref{tab:scan}, Fig.~\ref{fig:scan}).

\paragraph{Two bases, and what the gates said.} On the relaxed cell the $d$-only model of the
sealed verdict passes the filling gate cleanly ($\Delta \leq 0.004$, exact on the DFT mesh) but
fails the band-count criterion of R5 with the sealed frozen window ($\EF\pm1.5$~eV): at 8 of 1728
$k$-points near $(\pm0.42,\pm0.42,\pm0.25)$ a $d$ band that lies 13~meV below the window bottom in
DFT is placed 1.5~meV inside it by the model. This is a border effect of disentanglement, not a
missing orbital (the mean absolute error inside the window is 0.014~meV), and it is cured by
lowering the frozen bottom to $\EF-1.8$~eV (count exact, MAE 0.015~meV), but the cure is not
transferable: at $\hP = 1.33$~\AA{} the same model leaks at both windows. The $d+p$ model (16
functions, Fe-$d$ $+$ P-$p$), which had failed R5 in the first version of this paper because of an
interloper state from a too-wide outer window, passes both gates at every point once its outer
window is set from the DFT eigenvalues themselves (the envelope of bands 19--34 plus 0.1~eV; count
exact, MAE 0.013--0.015~meV, filling $\Delta \leq 0.016$). The scan therefore uses the $d+p$ model
as the production basis, with the Kanamori interaction on the ten Fe-$d$ orbitals only, exactly as
for SrRu$_2$As$_2$ in~\cite{ref1}; the $d$-only model is kept where it passes (the relaxed cell
and $\hP = 1.25$~\AA) as a basis cross-check.

\begin{figure}[htbp]\centering
\includegraphics[width=0.7\linewidth]{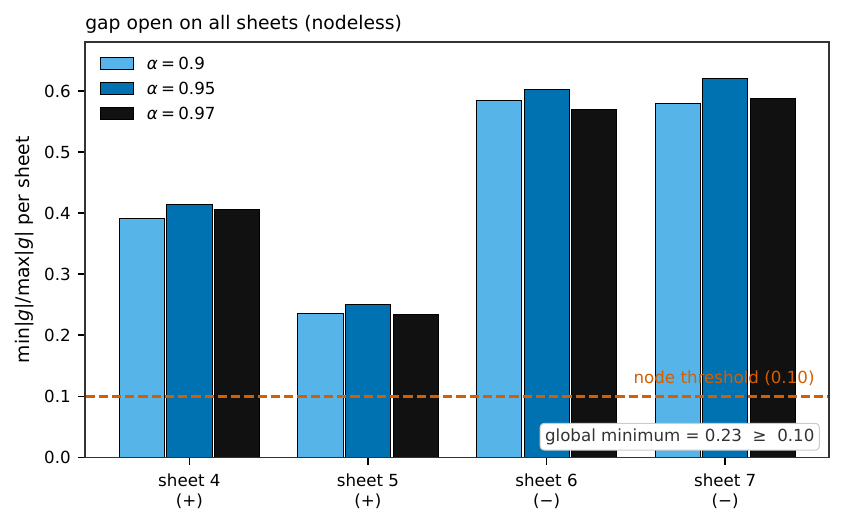}
\caption{Nodeless ratio $\min|g|/\max|g|$ per Fermi sheet at $\alpha=0.90/0.95/0.97$, relaxed cell,
$d$-only basis. Every sheet is $\geq0.23$, above the 0.10 node threshold, with the smallest margin
on the inner hole cylinder (sheet 5); in the $d+p$ basis the same sheet sits at 0.07--0.10
(Table~\ref{tab:scan}).}
\label{fig:nodeless}
\end{figure}

\paragraph{The relaxed cell.} With the $d$-only basis the relaxed cell returns
$\Ucrit = 1.68$~eV, $\lambda_1 = 0.068/0.122/0.166$ at $\alpha = 0.90/0.95/0.97$, the same nesting
vector $q^* = (\tfrac{1}{2},\tfrac{1}{2},\tfrac{1}{2})$, a clean inter-sheet sign reversal, and a nodeless
ratio on the weakest sheet of 0.23--0.25, up from 0.15--0.18 on the CHGNet cell
(Fig.~\ref{fig:nodeless}). Relaxing the cell thus moved the prediction \emph{away} from the nodal
line, not toward it, by the 0.02~\AA{} rise of the phosphorus and the 4\% contraction of the
stacking axis.

\paragraph{The explicit ligand changes the answer.} With the $d+p$ basis on the same relaxed cell
the sign reversal between hole and electron sheets survives intact (sign fractions 1/1/0/0,
$q^* = (\tfrac{1}{2},\tfrac{1}{2},\tfrac{1}{2})$, $A_{1g}$ projection 0.06--0.08 against $B_{2g} = 0$), but
the gap on the inner hole cylinder collapses to $\min|g|/\max|g| = 0.10/0.08/0.07$ at
$\alpha = 0.90/0.95/0.97$, at and below the reader's node threshold, and the reader returns
``\spm{}, near-nodal''. On the CHGNet cell the same basis gives 0.05/0.03/0.02. The other sheets
in the same basis are at 0.12--0.13 (outer hole cylinder) and 0.26--0.37 (electron sheets), so the
inner hole sheet carries a gap three to five times smaller than the electron sheets, an ordering
that does not depend on the reader's threshold. The threshold does decide the label: at 0.15 the
sealed $d$-only verdict (0.15--0.18) would sit on the line and the relaxed $d$-only one
(0.23--0.25) would stay nodeless; at 0.05 the relaxed $d+p$ result (0.07--0.10) would read nodeless
with a small gap, and the CHGNet $d+p$ result (0.02--0.05) would remain at the line. The critical
interaction also moves, from 1.68 to 2.37~eV, because part of the spectral weight that the $d$-only
model attributes to iron now sits on phosphorus and the Kanamori vertex, applied to the ten Fe-$d$
orbitals only, sees a weaker projected susceptibility; this is the cross-basis caveat of
Sec.~\ref{sec:cal} made concrete, and 2.37~eV is to be compared with the 4.17~eV of the
$d+p$ ruthenide rather than with the 0.97~eV of the $d$-only BaFe$_2$As$_2$. The difference between
the two bases is physical, not numerical: both reproduce the DFT bands to 0.015~meV and the DFT
occupations to $\Delta \leq 0.016$; what the $d$-only model cannot represent is the phosphorus weight
on the inner hole sheet, and it is on that sheet that the gap is decided.

\paragraph{Along the height.} Raising the phosphorus reproduces, inside one compound and one
instrument, the two ingredients of the Kuroki mechanism~\cite{kuroki2009}. At $\hP = 1.25$~\AA{} a
third hole pocket appears at the zone corner ($d_{xy}$; the Fermi surface goes from four to five
sheets), the nesting vector switches from $(\tfrac{1}{2},\tfrac{1}{2},\tfrac{1}{2})$ to the stripe vector
$(\tfrac{1}{2},\tfrac{1}{2},0)$ of BaFe$_2$As$_2$, and \Ucrit{} falls monotonically (2.37 $\to$ 2.04 $\to$
1.50 $\to$ 1.37~eV in the $d+p$ basis; 1.68 $\to$ 1.40 in the $d$-only), i.e.\ the system moves toward
the Stoner instability as the anion rises, the trend behind the empirical $T_c$--\hP{} curve. At the
same time the leading pairing channel loses its identity: $\lambda_2/\lambda_1$ climbs from 0.40 at
the physical height to 0.77, 0.88 and 0.95, and from $\hP = 1.33$~\AA{} the eigenvector that comes
out first changes sign inside every sheet (sign fractions $\approx0.5$, nodes on all sheets), the
signature of a $d$-wave-like solution nearly degenerate with the \spm{} one. The near-degeneracy
of \spm{} and $d$ channels in five-orbital RPA is known~\cite{graser}; here it is displayed as a
function of the one structural parameter, with the $d$-only basis, where it passes the gates,
staying nodeless \spm{} throughout. The reader labels the 1.33~\AA{} point ``$d$'' and the 1.42~\AA{}
point ``null'' (its degeneracy gate, $\lambda_2/\lambda_1 > 0.9$, fires); we report both labels as
what they are, an instrument declaring that it cannot name a single class there.

\paragraph{What the prediction becomes.} The sealed verdict survives in its sign structure and in
its wave vector, and fails in its nodeless character once the ligand is explicit: NaFeP at its
physical height is, by this instrument, an \spm{} superconductor with a gap that is nodeless on
three sheets and at the node threshold on the inner hole cylinder. That is the LiFeP side of the
height rule, not the LiFeAs side, and it is what the empirical rule of Sec.~\ref{sec:height}
predicted. The official prediction of this paper is therefore the $d+p$ reading, weaker and more
specific than the sealed one: inter-sheet sign reversal, a near-nodal inner hole sheet, and a
low-$T_c$ phosphide. The
manufacturability profile of Sec.~\ref{sec:funnel} is conditional on the nodeless reading and must
be read with that caveat: accidental near-nodes on one sheet are not symmetry nodes and do not by
themselves put a material in the coated-conductor class, but they erode the grain-boundary argument
that class~B rests on.
\begin{table}[htbp]\centering\small
\begin{tabular}{llcccccl}
\toprule
cell / point (\hP, \AA) & basis & \Ucrit{} (eV) & $q^*$ & $\lambda_1$ & $\lambda_2/\lambda_1$ & weakest-sheet $\min|g|/\max|g|$ & verdict\\
\midrule
CHGNet cell (1.14), sealed & $d$ & 1.72 & $(\tfrac{1}{2},\tfrac{1}{2},\tfrac{1}{2})$ & 0.108 & 0.30 & 0.18 (0.15--0.18) & \spm{} nodeless\\
CHGNet cell (1.14) & $d+p$ & 2.47 & $(\tfrac{1}{2},\tfrac{1}{2},\tfrac{1}{2})$ & 0.068 & 0.44 & 0.03 (0.02--0.05) & \spm{}, near-nodal\\
relaxed cell (1.16) & $d$ & 1.68 & $(\tfrac{1}{2},\tfrac{1}{2},\tfrac{1}{2})$ & 0.122 & 0.24 & 0.25 (0.23--0.25) & \spm{} nodeless\\
relaxed cell (1.16) & $d+p$ & 2.37 & $(\tfrac{1}{2},\tfrac{1}{2},\tfrac{1}{2})$ & 0.069 & 0.40 & 0.08 (0.07--0.10) & \spm{}, near-nodal\\
$\hP = 1.25$, 5 sheets & $d$ & 1.40 & $(\tfrac{1}{2},\tfrac{1}{2},0)$ & 0.208 & 0.40 & 0.22 (0.18--0.24) & \spm{} nodeless\\
$\hP = 1.25$, 5 sheets & $d+p$ & 2.04 & $(\tfrac{1}{2},\tfrac{1}{2},0)$ & 0.118 & 0.77 & 0.11 (0.04--0.15) & \spm{}, near-nodal\\
$\hP = 1.33$, 5 sheets & $d+p$ & 1.50 & $(\tfrac{1}{2},\tfrac{1}{2},0)$ & 0.076 & 0.88 & 0.00 (nodes within sheets) & $d$-like; \spm{}/$d$ degenerate\\
$\hP = 1.42$, 5 sheets & $d+p$ & 1.37 & $(\tfrac{1}{2},\tfrac{1}{2},0)$ & 0.069 & 0.95 & 0.00 (nodes within sheets) & degenerate (reader: null)\\
\bottomrule
\end{tabular}
\caption{The verdict along the pnictogen-height axis, in the two bases that pass the gates. All
rows: $24^3$ mesh, complete \chio{}, Kanamori blocks of five on the Fe-$d$ orbitals, $J/U = 0.15$,
$T = 130$~K; $\lambda_1$ and $\lambda_2/\lambda_1$ at $\alpha = 0.95$; the nodeless ratio is that of the
weakest sheet at $\alpha = 0.95$ with its range over $\alpha = 0.90$--$0.97$ in parentheses; verdict by the
frame-independent reader with the Stoner gate. The $d$-only model passes the gates only at
$\hP \leq 1.25$~\AA{} (Sec.~\ref{sec:scan}).}
\label{tab:scan}
\end{table}

\section[Manufacturability profile (conditional on the nodeless reading)]{Manufacturability profile (conditional on the nodeless reading)}\label{sec:funnel}
\begin{figure}[htbp]\centering
\includegraphics[width=0.95\linewidth]{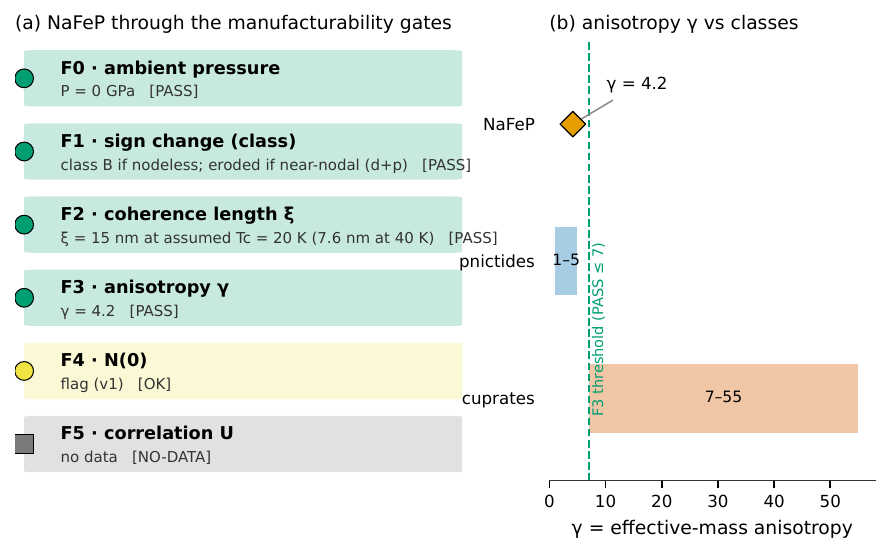}
\caption{(a) NaFeP through the manufacturability gates F0--F5; the coherence lengths are at assumed
$T_c$ values, since no $T_c$ is predicted; F1 is conditional on the nodeless reading. (b)
Effective-mass anisotropy $\gamma$: NaFeP (4.2)
against the pnictide (1--5) and cuprate (7--55) ranges; the F3 pass threshold is $\gamma\leq7$.}
\label{fig:funnel}
\end{figure}

The funnel gates below (Fig.~\ref{fig:funnel}; definitions and thresholds in~\cite{ref1}) are
consequences of the \spm{} class under its nodeless reading, not independent confirmations of it;
under the official near-nodal reading the class-B entry is eroded (F1 of Table~\ref{tab:funnel}). The
funnel does not predict $T_c$; where a gate needs one, it is stated as an assumption. Conditional on
that prediction, NaFeP has the following wire-relevant profile:
\begin{table}[htbp]\centering\small
\begin{tabularx}{\linewidth}{lX}
\toprule
gate & result\\
\midrule
F0 stability & ambient (0~GPa), $e_{\mathrm{hull}}\approx0$ (CHGNet and OQMD) --- pass\\
F1 grain-boundary class & class~B under the nodeless reading (\spm{}, grain-boundary tolerant, round wire viable); eroded under the near-nodal reading of the $d+p$ basis\\
F2 coherence length $\xi_0$ & 15~nm at an \emph{assumed} $T_c = 20$~K (17--51~nm over the 6--18~K range of the 111 family); worst sheet, $v_F = 2.2\times10^5$~m/s --- pass\\
F3 anisotropy $\gamma$ & 4.2 ($\leq7$) --- pass, nearly isotropic (3.6 on the CHGNet cell: the shorter relaxed $c$ axis raised it by 0.6)\\
F4 $N(\EF)$ & 2.3 states/eV per cell (2 Fe; Alexandria PBE 2.18~\cite{alexandria}), 11.6 states/(eV\,nm$^3$) --- non-blocking flag in this version\\
F5 Mott distance $U/W$ & not evaluated ($U$ not passed to the funnel) --- no data\\
\bottomrule
\end{tabularx}
\caption{Wire funnel for NaFeP, conditional on the predicted symmetry. $\xi_0 = \hbar\langle v_F\rangle/(\pi\Delta_0)$ with $\Delta_0 = 1.76\,k_BT_c$ on the worst Fermi sheet; $\gamma = \sqrt{\langle v_{ab}^2\rangle/\langle v_c^2\rangle}$ weighted over the Fermi surface; thresholds as pre-registered in~\cite{ref1}.}
\label{tab:funnel}
\end{table}

The anisotropy $\gamma = 4.2$ is far more isotropic than the cuprates ($\gamma\approx7$--55, which
require coated-conductor architectures) and in the range of the iron pnictides (1--5). Conditional
on the nodeless reading, NaFeP would be a fabrication-tolerant, arsenic-free, stoichiometric,
ambient-pressure wire candidate; under the near-nodal reading of the $d+p$ basis the class-B
argument weakens, since a gap that nearly closes on one sheet is more sensitive to grain-boundary
scattering than a fully open one, although accidental near-nodes are not the symmetry nodes that
force a coated conductor. The chemical risks flagged for synthesis are the large Na$^+$ ionic
radius relative to the FeP layer and the near-degenerate $Cmma$ polymorph of Sec.~\ref{sec:select}.

\section{What the prediction is, what it is not, and how to refute it}\label{sec:what}
The pipeline delivers, for a compound never made as a bulk phase: a crystallographic synthesis
target (NaFeP, $P4/nmm$; $a = 3.79$~\AA, $c = 6.68$--$6.78$~\AA{} across three independent PBE
relaxations, Sec.~\ref{sec:methods}), a paramagnetic DFT ground state, a predicted \spm{} gap
symmetry, near-nodal on the inner hole sheet in the $d+p$ basis that is the official prediction
and nodeless in the $d$-only basis of the sealed verdict, with the calibration error basis of
Sec.~\ref{sec:cal} and the
height scan of Sec.~\ref{sec:scan}, and a manufacturability prior conditional on the nodeless
reading. It does
\textbf{not} deliver a $T_c$ in Kelvin---the calibration establishes symmetry class, not absolute
$T_c$---nor does it substitute for the bench. By analogy to the 111 family and by the anion-height
systematics~\cite{mizuguchi}, a phosphide at $\hP\approx1.15$~\AA{} sits at the low-$T_c$ end of
the family (LiFeP 6~K, LaFePO 4--7~K), not at the LiFeAs end, and the moderate \Ucrit{} (1.68~eV
in the $d$-only basis, 2.37~eV in the $d+p$ basis, each farther from the Stoner point than
BaFe$_2$As$_2$ on its own scale) points the same way; but this is analogy, not
calculation, and no number is asserted.

The prediction is refuted or confirmed by the following measurements, in order of cost: (i)
synthesis (solid-state or flux; the Na$^+$-radius and polymorph risks) and refinement of the
structure---a $P4/nmm$ 111 phase with \hP{} in the range above is the target, and a refined \hP{} is
itself a test of the relaxed cells; (ii) susceptibility and transport for magnetic order and for a
superconducting transition; (iii) the nodal/nodeless question by the same probes that settled LiFeP
against LiFeAs---penetration depth, thermal conductivity, NMR relaxation---which is the single
measurement that discriminates our verdict from the empirical rule of Sec.~\ref{sec:height}; (iv)
the sign change between hole and electron sheets by quasiparticle interference or a neutron
resonance. A nodal or near-nodal gap confirms the $d+p$ reading and the height rule and refutes
the sealed $d$-only verdict; a fully open gap on all sheets does the reverse; no superconductivity
with no magnetic order refutes the Stoner-proximity reading of \Ucrit{} in both bases; a
stripe-ordered ground state refutes the paramagnetic input. Every field the pipeline does not
compute (synthesis route, kinetics, disorder) is left explicitly to the experimental partner in the
dossier.

\section{Cost and reproducibility}\label{sec:cost}
From the unreported compound to the sealed gap-symmetry verdict with a manufacturability profile
took under four hours of wall-clock time and about US\$\,3 of compute (SCF/NSCF and projections on a
spot A100; the Wannier model on an on-demand high-CPU node; \chio{} and the gap equation on a Colab
GPU). The robustness campaign of Sec.~\ref{sec:scan}---relaxation, four magnetic configurations on
two cells, four heights with NSCF, two Wannier bases at every height, and seven complete
\chio{}-plus-gap solutions---took one night and about US\$\,20 more (1.8~h of spot A100, 2~h of a
88-vCPU node, 3~h of Colab GPU). The candidate was itself the survivor of a zero-cost
literature-and-stability screen that eliminated 57 of 58 analogues. As in~\cite{ref1}, every result
file embeds its complete configuration (Fermi energy, interaction parameters, both meshes, energy
window, the hash of the tight-binding model, code version), so any verdict can be re-run; the \chio{}
recomputation reported in Sec.~\ref{sec:methods} reproduced the sealed result bit-for-bit.

\paragraph{The sealed record.} The verdict of Table~\ref{tab:gap} was sealed on 14 September 2026 in
the result file \texttt{gap\_nafep\_0914.json}, whose embedded configuration records the Fermi energy
(8.7463~eV), both meshes ($24^3$, $12^3$), $T = 130$~K, $J/U = 0.15$, the Kanamori block structure,
the absence of an energy window, and the SHA-256 hashes of the tight-binding model
(\texttt{e87aea14\ldots}) and of the Fermi-surface sample (\texttt{ce87902f\ldots}). The file
itself (SHA-256 \texttt{67cff0e9\ldots}) is provided as an ancillary file of this arXiv
submission, so that any future measurement on NaFeP can be compared against a prediction that
was dated before the measurement; the tight-binding model, the susceptibility checkpoint and the
gap eigenvectors are kept in the project's object store and are available from the author on
request. The Appendix
(Sec.~\ref{sec:app}) lists every database consulted for NaFeP, with query and date, and the
chronology of 14--15 September 2026 against the first NaFeP calculation. The list of the 58
screened compositions with the reason each was rejected and the sealed result file are provided
as ancillary files of this arXiv submission (\texttt{anc/}); the dated pre-registration of the
robustness campaign and the configuration blocks and gate reports of its seven runs are available
from the author on request, as in~\cite{ref1}.

\section{Conclusion}
We have applied the judge of~\cite{ref1} to NaFeP, an iron phosphide with no bulk synthesis and no
band-resolved electronic-structure study that survived the judge's own mechanism-and-stability
screen, and
predict it to be an ambient-pressure, arsenic-free, paramagnetic \spm{} superconductor. The sealed
first verdict called the gap nodeless; the robustness work that followed---a PBE relaxation, the
stripe and N\'eel tests, a second Wannier basis with the ligand explicit, and a scan of the
phosphorus height---kept the sign reversal and the nesting vector and moved the inner hole sheet to
the node threshold, where the empirical pnictogen-height rule had placed it from the start. We
report both bases, adopt the ligand-explicit one as the official prediction, and report the
direction in which the physics moved rather than the more attractive of the two numbers. This is the first gap-symmetry prediction that exists for NaFeP, offered as a
falsifiable synthesis target with the instrument's calibration error basis stated and no $T_c$
asserted, and the first time this instrument has been run against a structural rule inside a
single compound. It is the concrete next step after the conclusion of~\cite{ref1}---that generation
re-finds the known and judgment is the bottleneck: here the judgment is applied, it names a
specific candidate for the bench, and it names the measurement that decides between its own two
readings. We invite experimental collaboration to synthesize NaFeP and measure the gap on its
inner hole sheet.

\section{Appendix: databases consulted, and the chronology of the decisions}\label{sec:app}
Table~\ref{tab:s1} records every database consulted for NaFeP, with the query and the date, and
what each returned; Table~\ref{tab:s2} places every decision of 14--15 September 2026 that
affects the reader or the gates against the first NaFeP calculation. Local time is UTC$-3$; the
object-store timestamps behind the table are UTC.

\begin{table}[htbp]\centering\scriptsize
\begin{tabularx}{\linewidth}{lllX}
\toprule
database & date (2026) & query & result\\
\midrule
Materials Project~\cite{mp} & 14 Sep; 15 Sep (OPTIMADE) & Na--Fe--P, $n_{\mathrm{el}} = 3$ & no Na--Fe--P ternary\\
OQMD~\cite{oqmd} & 14 Sep (entry page); 15 Sep API: HTTP 502 & NaFeP & entry 1370778, $P4/nmm$, on hull, $-0.61$~eV/atom; $Cmma$ $+0.4$~meV/atom; no ICSD id; band gap 0 only\\
Alexandria, PBE~\cite{alexandria} & 15 Sep (OPTIMADE) & reduced formula FeNaP & agm002166081, $P4/nmm$, hull 0.000, $-0.480$~eV/atom, $a = 3.797$, $c = 6.698$, $\hP = 1.150$~\AA; gap 0, $N(\EF) = 2.18$ states/eV per cell, moment 0; no band structure; four other polymorphs 0.16--0.24~eV/atom above the hull\\
Alexandria, PBEsol~\cite{alexandria} & 15 Sep (OPTIMADE) & same & same entry, hull 0.000, $-0.515$~eV/atom, $a = 3.748$, $c = 6.653$, $\hP = 1.137$~\AA; gap 0, $N(\EF) = 2.47$, moment 0\\
JARVIS-DFT~\cite{jarvis} & 15 Sep (OPTIMADE) & FeNaP; Na--Fe--P & 0 entries\\
AFLOW~\cite{aflow} & 15 Sep (AFLUX) & species Fe, Na, P & 104 prototype decorations; the two $P4/nmm$ tP6 entries have Na on the 2a square-net site and Fe on 2c ($+0.108$~eV/atom, $0.57\,\mu_B$/atom): anti-site decorations, not the 111 compound; ICSD catalogue 0\\
COD~\cite{cod} & 15 Sep (OPTIMADE) & Na--Fe--P, $n_{\mathrm{el}} = 3$ & 0 entries\\
ICSD & not accessible (licensed) & --- & indirect only: COD 0, AFLOW ICSD catalogue 0, literature gate of 14 Sep 0\\
Literature (web, arXiv) & 14 Sep; 15 Sep & ``NaFeP'' phosphide, superconductivity, DFT & only the battery mention of~\cite{yan2025}; no band-structure, Fermi-surface, magnetic or pairing study\\
\bottomrule
\end{tabularx}
\caption{Every database consulted for NaFeP. Alexandria holds a PBE and a PBEsol calculation
with a metallic gap, a density of states at \EF{} and a zero moment, and no band structure; this
is why the paper claims the first band-resolved electronic structure and the first pairing
prediction, not the first electronic-structure calculation.}
\label{tab:s1}
\end{table}

\begin{table}[htbp]\centering\scriptsize
\begin{tabularx}{\linewidth}{lXl}
\toprule
local time (14 Sep 2026) & event & evidence\\
\midrule
08:17--08:32 & literature and stability gate on the 58 Fe analogues; NaFeP the sole survivor & screen files\\
09:05 & pre-registration of the NaFeP run (kill criteria, three-line prediction) & manifest\\
\textbf{09:25} & \textbf{first NaFeP calculation}: non-magnetic SCF; 09:28 ferromagnetic start collapses & SCF outputs\\
10:36--10:50 & clean SCF, NSCF on $12^3$, projections; $\EF = 8.7463$~eV, Fe-$d$ 80\% & NSCF outputs\\
11:17--12:12 & Wannier B ($d$-only) and A ($d+p$, first build; fails the R5 count) & Wannier logs\\
\textbf{12:52} & \textbf{sealed gap verdict}: \spm{} nodeless, $\Ucrit = 1.72$~eV, inner sheet 0.15--0.18 & \texttt{gap\_nafep\_0914.json}\\
13:46 & LSCO forensic check: the solver's verdict string found frame-blind & record\\
\textbf{14:16} & \textbf{Nb kill-test sealed}; $\Ucrit = 3.65$~eV from the checkpoint & \texttt{gap\_nb\_0914.json}\\
$\approx$14:30--15:01 & frame-independent reader written; node threshold 0.10 fixed from the sealed suite (NaFeP 0.15--0.18 already known) & reader source, 15:01\\
\textbf{14:52} & Nb labelled ``$d$'' by the shape-only reader; \textbf{Stoner gate-0 at 2.5~eV proposed} & commit\\
\textbf{15:01} & \textbf{2.5~eV coded}; the source comment lists NaFeP 1.72 among the values it separates & reader source\\
18:22--22:08 & first revision of this paper; 2.5~eV re-described as the YBCO--Nb midpoint & revision files\\
22:04 & pre-registration of the robustness campaign (relaxation, N\'eel and stripe orders, height scan, rebuilt $d+p$); machines launched after it & campaign manifest\\
22:04--02:30 (15 Sep) & relaxation, magnetic orders, four heights, both bases with R5 and filling gates, seven \chio{}-plus-gap solutions; $d+p$ puts the inner hole sheet at 0.07--0.10 & campaign outputs\\
\bottomrule
\end{tabularx}
\caption{Chronology of 14--15 September 2026. Both reader thresholds (the node ratio 0.10 and the
Stoner gate 2.5~eV) were fixed after the sealed NaFeP verdict and with its values in hand; the
official $d+p$ prediction comes from a campaign pre-registered after the sealed verdict and run
after that pre-registration.}
\label{tab:s2}
\end{table}

\end{document}